\documentclass{aa}

\usepackage{graphicx}
\begin{document}

\title{4\,$\mu$m spectra of AGB stars I. Observations}

\author{T.~Lebzelter\inst{1}
\and K.H.~Hinkle\inst{2}
\and B.~Aringer\inst{1}}

\offprints{lebzelter@astro.univie.ac.at}

\institute{Institut f\"ur Astronomie, Universit\"at Wien, 
T\"urkenschanzstr.\,17, A-1180 Vienna, Austria
\and
Kitt Peak National Observatory, 
National Optical Astronomy Observatory
\thanks{Operated by the Association of Universities for Research
in Astronomy, Inc.~under cooperative agreement with the National Science
Foundation},
950 N.Cherry Avenue, P.O.Box 26732, Tucson, Arizona 85726
}

\date{Received / Accepted}

\abstract{
We present times series of high resolution spectra of AGB variables at
4\,$\mu$m. Line profiles from the major contributors to the spectra of oxygen 
rich stars
at 4\,$\mu$m, OH, H$_2$O, HCl and SiO, are examined.  The 
velocity as well as shape variations of these profiles 
with time are discussed.  
The line profiles investigated frequently have emission and multiple
absorption components. 
The changes with time of the 4\,$\mu$m
region lines
do not always follow the cyclic variability seen in NIR spectra and in the 
photometric light curve.  
We interpret and discuss the
results qualitatively considering comparing the spectral variability with
that of the well behaved 1.6\,$\mu$m region and 
of dynamical model atmospheres.  Miras
and semiregular variables are compared. 
The origins of non-periodic behavior are discussed, 
including the role of spatial inhomogeneities in the stellar atmosphere. 
}

\maketitle

\section{Introduction}

The structure of the envelopes of Asymptotic Giant Branch (AGB) stars is
dominated by large amplitude stellar pulsations.  The mechanical energy
of the pulsation can be larger than L/c by orders of magnitude (Hinkle
et al.~\cite{HHR}), resulting in a cool, highly extended outer
atmosphere.  As a result of the pulsational flow of mechanical energy,
the structure of the atmosphere varies strongly during the light
cycle.  It is possible with time series spectra to monitor the flow of
material in the stellar atmosphere.  In addition, velocities measured
from different spectral transitions of different atomic and molecular species
probe the dynamics at different (optical) depths in the atmosphere.

One result of the extended, cool atmosphere of these stars is that the
stellar flux is primarily emitted in the infrared.   Due to the rich 
infrared
molecular spectra of the cool AGB stars, the infrared is also the
optimum spectral region for studying the outer atmosphere of these
stars.  The analysis of time series of high resolution infrared spectra
has been used over three decades by Hinkle and collaborators
(e.g.~Hinkle et al.~\cite{HHR}, \cite{HSH84}, \cite{HLS97}). Both
miras and the lower amplitude semiregular variables have been 
investigated. The authors
mainly used CO lines in the 1.6-2.5\,$\mu$m region, since 
those transitions can
be easily identified and observed.  The high excitation first overtone
CO lines and the second overtone lines of CO show velocity variations
of 25 to 30\,km\,s$^{-1}$ in miras and about 10\,km\,s$^{-1}$ and 
less in semiregular
variables.  In the 1.5-2.5\,$\mu$m region other atomic and molecular
lines have been found to have similar behavior, while low excitation
first overtone lines of CO show only very small velocity changes with
phase.  It has been suggested that these spectral lines as well as some
of the H$_2$O lines probe a more extended pseudo-stationary layer.
Spectral lines with excitation potentials near zero volts and large
column densities, e.g. the lowest excitation CO  
fundamental lines situated around 4.6\,$\mu$m, 
probe the classic, expanding circumstellar shell (Ryde et al.~\cite{R00}).

In this paper we will investigate the molecular spectra in the 4\,$\mu$m
region.  In oxygen rich atmospheres, the principal molecular constituents 
are the first overtone SiO bands, the fundamental bands of OH, and
the HCl fundamental.  In addition, the 4\,$\mu$m region is blanketed by
a large number of weak water lines in stars with O$>>$C.  We will
discuss high spectral resolution observations of lines from all four 
of these molecules in this paper.  
This is the first use of these 4\,$\mu$m lines at 
high resolution for monitoring the atmospheric changes in pulsating
stars.  However, the 4\,$\mu$m spectrum of SiO has been monitored 
previously at medium
and low resolution (Hinkle et al.~\cite{HBLB76}; Rinsland \& Wing
\cite{RW82}).  

Aringer et al.~(\cite{AJL97}, \cite{AHWHJKW99}) have shown from
medium resolution 4\,$\mu$m spectra that using hydrostatic models
cannot reproduce the spectra properly.  In order to match the
observations one needs to introduce additional features like the ``warm
molecular envelope'' proposed by Tsuji et al.~(\cite{TOAY97}). On the
other hand, calculations based on recently developed dynamical
atmospheric models (H\"ofner et al.~\cite{HJLA98}) have the ability to
reproduce the observed spectral features as shown by Aringer et
al.~(\cite{AHWHJKW99}) for the SiO first overtone band heads around
4\,$\mu$m.  In this paper we will present the
observational data for a sample of miras and semiregular variables.
These data will provide the input for a forthcoming second paper
(Aringer et al.~in preparation) that will deal with a comparison of our
findings with atmospheric models focusing on the behavior of the SiO
lines as in the previous papers by Aringer et al.~(\cite{AJL97},
\cite{AHWHJKW99}).

\section{Molecular data}

\subsection{OH}

The hydroxyl free radical is a well known contributor of 
vibration-rotation lines to the spectra of K and M stars.
Aringer (\cite{Aringer00}) finds that 
OH appears in giant spectra at an effective temperature below approximately
4200\,K.  
The OH molecule is well known for the three maser lines in the 
rotational spectra of AGB stars at 1612, 1665
and 1667\,MHz, respectively.  These lines provide an important tool for
deriving accurate center-of-mass velocities of AGB stars (see Habing
\cite{Habing96} for a review).  Within our sample of stars OH masers have been
detected only in R\,Cas and R\,Leo (Benson et al.~\cite{BLW90}).
The maser lines are much lower energy transitions than
any of the OH lines discussed in this paper and are formed in the 
expanding circumstellar shell.

The 3-4\,$\mu$m region contains the vibration-rotation fundamental 
bands of OH. 
The first overtone vibration-rotation bands are found around 1.5\,$\mu$m.  
Due to the low molecular weight of OH, the rotational constant is 
large and OH does not form band heads in the P branches in the near infrared.  
The OH lines are out of the X$^2\Pi$ ground state and exhibit large
lambda doubling.  Hence the lines appear in quartets in the spectrum.
A detailed study of the first overtone OH lines 
in the mira variable R\,Leo is described by Hinkle (\cite{Hinkle78}). 
Line positions derived from Coxon (\cite{C80}) and Coxon \& Foster
(\cite{CF82}) are used in the current paper.  The line list consists
of a group of 17 1-0 and 2-1 P branch lines selected from the total
of 46 lines in the 2500 to
2800\,cm$^{-1}$ region.  The chosen lines have minimal 
blending with telluric and other stellar features.  The entire set
of 46 lines was used to evaluate the measurement process as described
below.

\subsection{SiO}

SiO is, after CO and H$_{2}$, one of the most abundant molecules
in the atmospheres of oxygen-rich AGB stars.  It is formed in all
cooler oxygen rich stars.  Model spectra indicate that significant amounts of
SiO are formed in
stellar atmospheres with effective temperatures of approximately
4000\,K or less
(Aringer \cite{Aringer00}).  In addition to being abundant, SiO is a
primary constituent of oxygen-rich dust.  Since dust is one of the
requirements for AGB mass loss (Gail \& Sedlmayr \cite{GS86}), SiO
formation is a prerequisite for mass loss.

SiO maser emission is very common among miras (Jewell et
al.~\cite{JSWWG91}).  Beside sharp maser lines, broad thermal SiO
emission in rotational lines 
has been detected in several AGB stars (see Habing
\cite{Habing96} for a review). Some SiO maser lines involve relatively large J
rotational transitions.  These lines could be formed near the same
region probed by the infrared vibration-rotation SiO lines.  Two
infrared SiO vibration-rotation transitions have been observed in
late-type stars.  The fundamental bands are found in the spectrum near
8\,$\mu$m.  The first overtone bands are found near 4\,$\mu$m.  We
report here on measurements of lines from the first overtone.

Unlike OH, the SiO line spacing is small.  This spacing combined with
the transitions occurring in a X$^1\Sigma$ state results in well
defined R-branch band heads.  The first high resolution spectra of the
SiO first overtone lines in a star were 
obtained by Beer et al.~(\cite {BLS74}) who
used the spectra to derive Si isotopic abundances in $\alpha$ Ori.
Observing the band heads at medium spectral resolution, Hinkle et
al.~(\cite{HBLB76}) found that long period variables show a strong
variability in the intensity of the SiO bands at 4\,$\mu$m. In miras,
around light maximum SiO becomes almost undetectable, strengthening
towards light minimum. This result has been confirmed by Rinsland \&
Wing (\cite{RW82}), who also noted that SiO absorption features are
most prominent in stars with the lowest temperatures (i.e.~at light
minimum in miras).  Different explanations for the strange behavior of
SiO in AGB variables have been discussed including dissociation,
additional absorption (e.g.~by H$_{2}$O) or emission of SiO (Rinsland
\& Wing \cite{RW82}, Aringer et al.~\cite{AHWHJKW99}).  The only
published high resolution spectra of the first overtone bands of SiO in
AGB variables is an atlas of the 4\,$\mu$m range by Ridgway et
al.~(\cite{RCHJ84}).

The SiO line list used in this paper is made up of 20 unblended 
$^{28}$SiO lines in the 2-0 R branch between
the 3-1 band head and the 2-0 band head (2-0 R22 to R46) spanning the
wavenumber interval 2475 to 2493\,cm$^{-1}$.  The frequencies used are
those reported in Hinkle et al.~(\cite{HWL95}).

\subsection{HCl}

There are no previous detailed investigations of HCl 
in cool evolved stars.  Ridgway
et al.~(\cite{RCHJ84}) report lines of HCl in the 4\,$\mu$m spectrum of
the S-type mira R\,And, but they did not find HCl in oxygen- or
carbon-rich AGB stars.  The HCl vibration-rotation fundamental 
crosses the 3-4\,$\mu$m
region.  The transition is a X$^1\Sigma$ and since HCl, like OH, is a
hydride the rotational constant is large and hence the line spacing is
large without prominent band
heads.  The frequencies of the lines are taken from 
Le Blanc et al.~(\cite{LWB94}).  Lines of both 
the H$^{35}$Cl and H$^{37}$Cl isotopes
were used.  This augments the rather spare number of lines present and
since the solar system $^{35}$Cl/$^{37}$Cl is small (about
three) the intensities of the lines from the two isotopes should be 
similar.  The line list includes 20 lines with the least blending selected 
from the \mbox{1-0}
and \mbox{2-1} bands limited to the wavenumber region between 2500 and
2800\,cm$^{-1}$.

\subsection{H$_{2}$O}

Water was first observed in the infrared spectra of cool stars in the
early 1960s (see Spinrad \& Wing \cite{SW69}).  Later in that decade
observations showed water bands to be very strong, variable features in
mira spectra (Johnson \& Mendez \cite{JM70}).  Recently Alvarez et
al.~(\cite{ALPW00}) concluded that the detection of deep water bands in
stellar spectra is a direct indication of pulsation.  Early workers
also recognized that despite water's large abundance in cool,
oxygen-rich stars, observations of this molecule from the ground are
inhibited by water bands in the earth atmosphere.  Water must either be
observed from space (Aringer et al.~\cite{AringerAGB99}; 
Matsuura et al.~\cite{MYMFT99}; Truong-Bach et al.~\cite{TB99}; Yamamura et
al.~\cite{YJC99}) or by using excited levels which are populated in
atmospheres of stars but are not populated in the Earth's atmosphere
(Hinkle \& Barnes \cite{HB79a}).

In addition to the vibration-rotation infrared spectrum of H$_{2}$O,
this molecule has a rich pure rotational spectrum in the microwave and
far-infrared.  While considerable blockage of this spectrum also occurs
as a result of water in the Earth's atmosphere, water maser lines are
known in circumstellar envelopes of evolved stars (Menten \& Melnick
\cite{MM89}) including three of our program stars, namely R\,Cas
(Colomer et al.~\cite{CRMB00}), R\,Leo (Menten \& Young \cite{MY95}),
and RX\,Boo (Colomer et al.~\cite{CRMB00}).

The spectrum of water in the atmospheres of LPVs originates in two
distinct regions of the star's atmosphere: Most of the lines are formed
in the circumstellar region closest to the star, while some lines
follow the photospheric behavior and therefore originate much deeper in
the atmosphere (Hinkle \& Barnes \cite{HB79a}; Matsuura et
al.~\cite{MYMFT99}).  Infrared H$_{2}$O absorption cannot originate far
out in the circumstellar envelope as this would not be consistent with
the mass loss rates found in these objects.  Tsuji et
al.~(\cite{TOAY97}) and Tsuji (\cite{T00}) argue that the non-photospheric water
lines indicate the existence of a warm molecular envelope.

In this paper we will discuss high excitation lines of the $\nu_3$
vibration-rotation band which are present in the 4\,$\mu$m spectrum.
The band origin is at 3756\,cm$^{-1}$.  The line list consists of 102
high excitation $\nu_3$ lines between 2500 and 2800\,cm$^{-1}$.  The
frequencies are from Zobov et al. (\cite{Z00}).  The line list used for
this paper involves moderately high rotational (J) quantum numbers
between 12 and 24 and should not include lines formed in the inner
circumstellar region.  However, we note that like the OH and HCl
transitions, the $\nu_3$ transition is a fundamental vibrational
transition.  Fundamental transitions are the most likely to be detected
in cool conditions and are also the preferred routes for radiation
resulting from recombination.

\section{Observations}

The spectra were obtained by one of us (K.H.H.) with the
Fourier Transform Spectrometer (FTS) operated at the Mayall 4m
telescope at Kitt Peak National Observatory.  A description of the
instrument can be found in Hall et al.~(\cite{HRBY79}). The FTS is no
longer available for use, having been closed in the mid-1990's.  The spectra
are an inhomogeneous set, having been observed for several 
projects.  However, all cover the 2400 to 2800\,cm$^{-1}$ region
required by the line lists. The resolution achieved is about
R$=$100000 except for very few observations at a somewhat lower 
resolution but none below 50000.
The observations were obtained mainly
between 1984 and 1986 with a few additional spectra observed earlier.
Three spectra of $\chi$\,Cyg were observed in 1994 and 1995, shortly
before the FTS on Kitt Peak was closed. All data have been reduced
as described in Hinkle et al.~(\cite{HHR}).

\begin{table}
\begin{flushleft} 
\caption[]{Observing log for the miras. The period in column\,3 is taken from the
GCVS.}
\label{sample1}
\begin{tabular}{lllccc} 
\hline 
GCVS   & Var.  & Period & JD          & Visual & S/N\\
name   & type  &        & 2440000+    & phase  &  \\
\hline
\object{R Cas}  & Mira  & 430    & 5986 & 0.54   & 53.7 \\
       &       &        & 6018 & 0.61   & 106.7 \\
       &       &        & 6108 & 0.82   & 27.7 \\
       &       &        & 6160 & 0.94   & 87.9 \\
       &       &        & 6192 & 0.01   & 47.7 \\
       &       &        & 6223 & 0.09   & 52.1 \\
       &       &        & 6243 & 0.13   & 35.1 \\
       &       &        & 6312 & 0.29   & 92.6 \\
       &       &        & 6454 & 0.62   & 44.0 \\
       &       &        & 6510 & 0.75   & 50.2 \\
       &       &        & 6570 & 0.89   & 58.4 \\
\hline
\object{o Cet}  & Mira  & 332    & 3437 & 0.89   & 243.4 \\
       &       &        & 4775 & 0.81   & 108.6 \\
       &       &        & 5986 & 0.45   & 63.2$^{1}$ \\
       &       &        & 6018 & 0.55   & 93.4 \\
       &       &        & 6095 & 0.78   & ...$^{2}$ \\
       &       &        & 6108 & 0.82   & 57.6 \\
       &       &        & 6312 & 0.43   & 66.6 \\
\hline
\object{$\chi$ Cyg} & Mira & 408 & 3437 & 0.17   & 135.9 \\
       &       &        & 5986 & 0.35   & 53.8 \\
       &       &        & 6018 & 0.43   & 88.0 \\
       &       &        & 6108 & 0.65   & 75.0 \\
       &       &        & 6160 & 0.77   & 86.3 \\
       &       &        & 6192 & 0.85   & 85.8 \\
       &       &        & 6223 & 0.93   & 65.2 \\
       &       &        & 6243 & 0.98   & 36.0 \\
       &       &        & 6312 & 0.15   & 71.4 \\
       &       &        & 6346 & 0.23   & 59.4 \\
       &       &        & 6427 & 0.43   & 22.0 \\
       &       &        & 6454 & 0.50   & 50.6 \\
       &       &        & 6510 & 0.63   & 64.4 \\
       &       &        & 6570 & 0.78   & 65.4 \\
       &       &        & 9323 & 0.62   & 89.4 \\
       &       &        & 9363 & 0.71   & 93.7 \\
       &       &        & 9450 & 0.93   & 88.0 \\
       &       &        & 9612 & 0.30   & 120.1 \\
       &       &        & 9798 & 0.75   & 66.3 \\
\hline
\object{R Leo}  & Mira  & 310    & 4650 & 0.60   & 329.2 \\
       &       &        & 5986 & 0.76   & 70.3 \\
       &       &        & 6164 & 0.31   & 54.6 \\
       &       &        & 6427 & 0.15   & 24.4 \\
       &       &        & 6511 & 0.41   & 74.6 \\
\hline
\end{tabular}

\begin{tiny}
$^{1}$ two spectra\\
$^{2}$ no reverse scan\\
\end{tiny}
\end{flushleft}
\end{table}

\begin{table}
\begin{flushleft} 
\caption[]{Observing log for the semiregular variables. The period in column\,3 is taken from the
GCVS. Phases are rough estimates from AAVSO light curves.}
\label{sample2}
\begin{tabular}{lllccc} 
\hline 
GCVS   & Var.  & Period & JD          & Visual & S/N\\
name   & type  &        & 2440000+    & phase  &  \\
\hline
\object{RZ Ari}  & SRb   & 30     & 7071 & --    &      \\
\object{RX Boo}  & SRb   & 340    & 6986 & 0.0:  &  \\
\object{W Cyg}   & SRb   & 131    & 6160 & 0.4:  & 42.8 \\
\object{RU Cyg}  & SRa   & 233    & 6160 & 0.2:  & 16.9 \\
\object{g Her}   & SRb   & 89     & 6986 & 0.9:  & ...$^{1}$\\
\object{R Lyr}   & SRb   & 46     & 6986 & --    &      \\
\object{SW Vir}  & SRb   & 150    & 6986 & --    &      \\
\hline
\end{tabular}

\begin{tiny}
$^{1}$ no reverse scan\\
\end{tiny}
\end{flushleft}
\end{table}

For four miras the data set includes time series of 4\,$\mu$m spectra,
allowing the
stellar pulsation to be monitored.  There are also single epoch
observations of a number of semiregular variables.  Together with some
fundamental parameters from the General Catalogue of Variable Stars
(GCVS, Kholopov et al.~\cite{GCVS4}), the observational details 
are given in Tables\,\ref{sample1} and \ref{sample2}. Phases are calculated from
the closest light maxima provided by the AAVSO (Mattei \cite{AAVSO}).
As can be seen from Table\,\ref{sample1} for three of the four miras
(R\,Cas, o\,Cet and $\chi$\,Cyg) observations from
different light cycles but at similar phases exist. This can be used
to investigate the cycle to cycle behavior of these stars (see below).
The dust enshrouded object IRC+10216, which
is a featureless, bright continuum source at 4\,$\mu$m, was also observed.
The IRC+10216 spectrum allows easy identification of telluric lines.

Seven stars of our sample (all miras, RU\,Cyg, W\,Cyg, g\,Her) have
previously been investigated using 1.6\,$\mu$m spectra (Hinkle et
al.~\cite{HHR}, \cite{HSH84}, \cite{HLS97} and 
Lebzelter~\cite{Lebzelter99}). These papers also give a summary of the
fundamental properties for these stars from the literature.
Three miras of our sample are 'typical' representatives for oxygen rich
disk miras with respect to period, amplitude, spectral type and mass
loss. Since they are among the brightest miras in the sky, these
objects are very well studied. The fourth mira is $\chi$\,Cyg, a S-type
star, the pulsational behavior of which has been studied in detail by
Hinkle et al.~(\cite{HHR}) and Wallerstein (\cite{Wallerstein85}).

The spectra are very blended
and it is typically not possible to find `clean' individual line profiles. 
Hence we have worked from average line profiles calculated from all lines
of the list.
Velocities have been determined from the deepest point in the 
average profile of the lines for each of the four molecules
investigated.  In many cases this velocity summarizes a blend so, 
in addition, the average profiles themselves are discussed below.
The line lists typical contain about 20 
lines for each molecule of similar intensity and excitation. 
For water, many more lines were available
and as can be seen in the average profiles the `noise' was significantly
reduced.  

The average profile 
is modified by blending lines around each of the individual
line positions.  Given the high level of blending of some spectra
it is remarkable that a reasonable average profile emerges.
In this light it is critical that the average line profiles not 
be over interpreted.  We feel that the general features, especially
those seen in time series and in the profiles of other molecules,
are real.  However, detailed features seen only in one profile 
could be artifacts of the averaging process.  In the case of OH 
we performed a test where average profiles were made up from two independent
line lists of 23 lines each.  There was remarkable agreement between 
the two profiles, indicating that essentially all the features are 
real.

The velocity of telluric lines was measured to check on the zero point
of the velocity scale.  This correction is typically done in FTS
spectra because the reference and stellar beams can have non-parallel
paths.  In each of the 4\,$\mu$m spectra we found the telluric velocities to be
distributed around zero with a full width at half maximum of this
distribution of about 0.1\,km~s$^{-1}$.  At longer wavelength alignment
errors become less significant and this seems confirmed by this data
set.  Applying the small, apparently random telluric correction would
increase the uncertainties and hence we have not done so.  In any case
the correction is less than or on the order of $\pm$0.2\,km~s$^{-1}$ 
in the measured velocities.  Further details on 4m
FTS data and FTS data reduction can be found in Hinkle et
al.~(\cite{HSH84}, \cite{HWL95}).

\section{Mira Results}

\begin{figure}
\resizebox{\hsize}{!}{\includegraphics{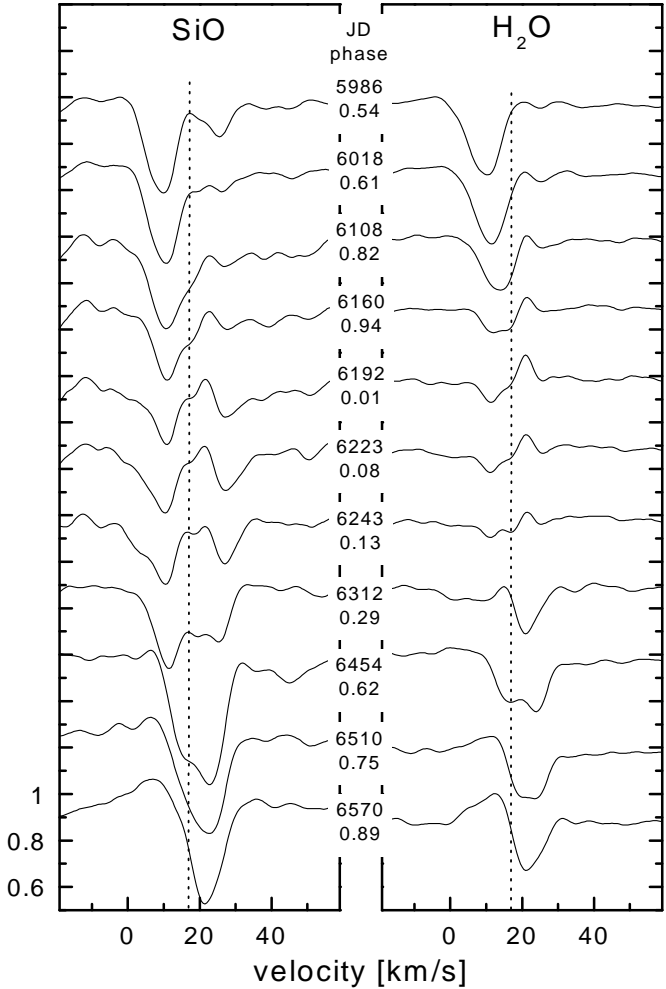}}
\caption[]{Averaged SiO 2-0 (left) and H$_{2}$O $\nu_3$ (right) 
line profiles of R Cas.
Date of observation and according phase are given in the
plot. The dotted line marks the center-of-mass velocity.}
\label{rcassioh2o}
\end{figure}

\begin{figure}
\resizebox{\hsize}{!}{\includegraphics{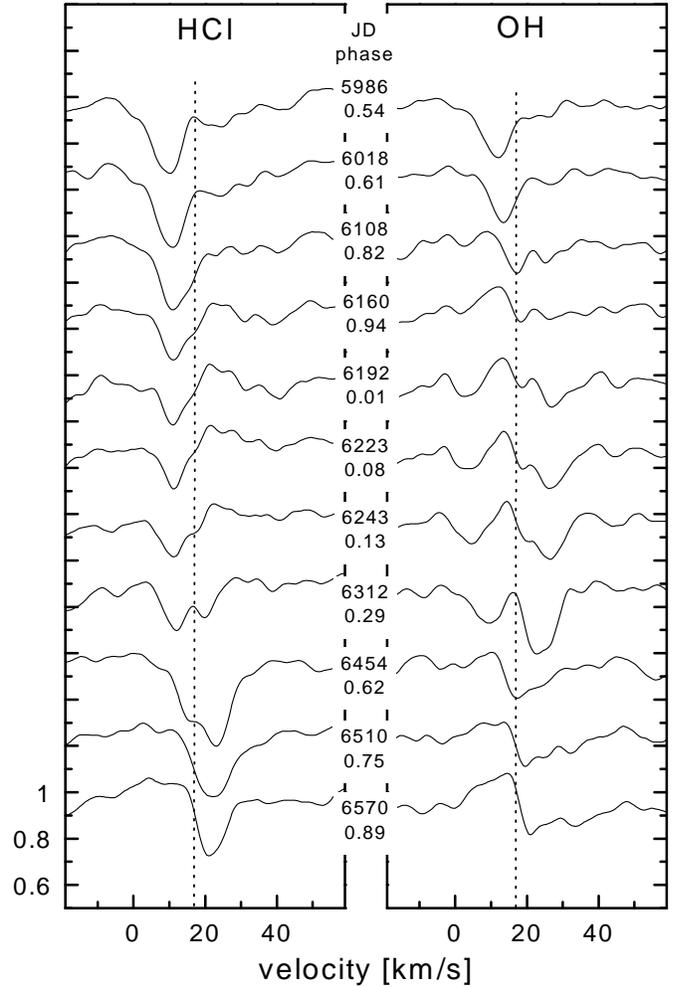}}
\caption[]{Averaged HCl fundamental 1-0, 2-1 (left) and OH 
fundamental 1-0, 2-1 (right) line profiles of R Cas.
Date of observation and according phase are given in the
plot. The dotted line marks the center-of-mass velocity.}
\label{rcashcloh}
\end{figure}

\begin{figure}
\resizebox{\hsize}{!}{\includegraphics{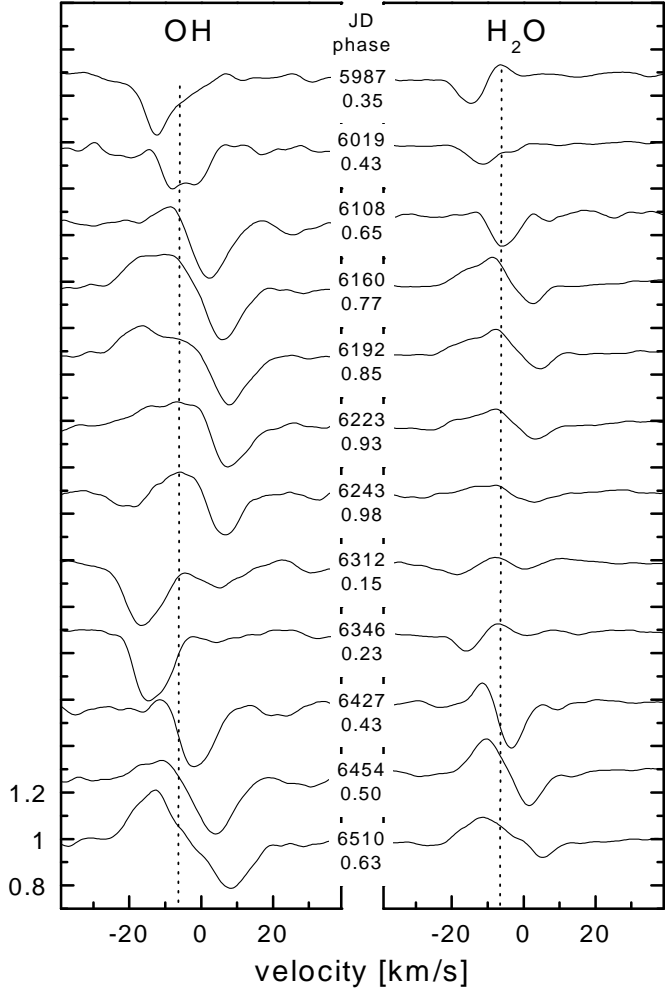}}
\caption[]{Averaged OH fundamental (1-0, 2-1) (left) and
H$_2$O $\nu_3$ (right) line profiles of $\chi$\,Cyg.
Date of observation and according phase are given in the
plot. The dotted line marks the center-of-mass velocity.}
\label{ccygohh2o}
\end{figure}

\begin{figure}
\begin{center}
\resizebox{\hsize}{!}{\includegraphics{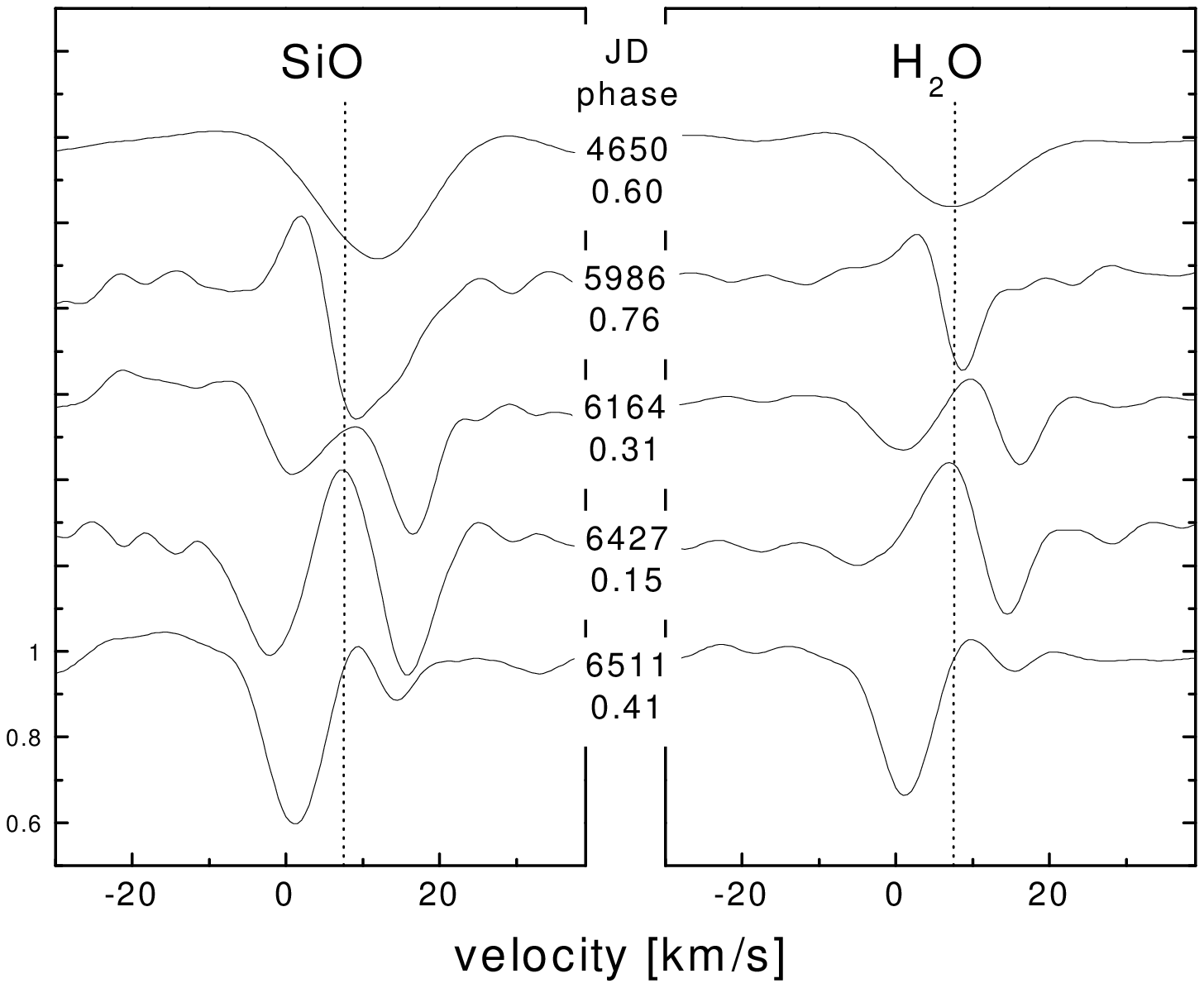}}
\end{center}
\caption{SiO \& H$_{2}$O line profiles of R Leo.
Same as Figure\,\ref{rcassioh2o}.}
\label{rleosioh2o}
\end{figure}

\begin{figure}
\begin{center}
\resizebox{6.5cm}{!}{\includegraphics{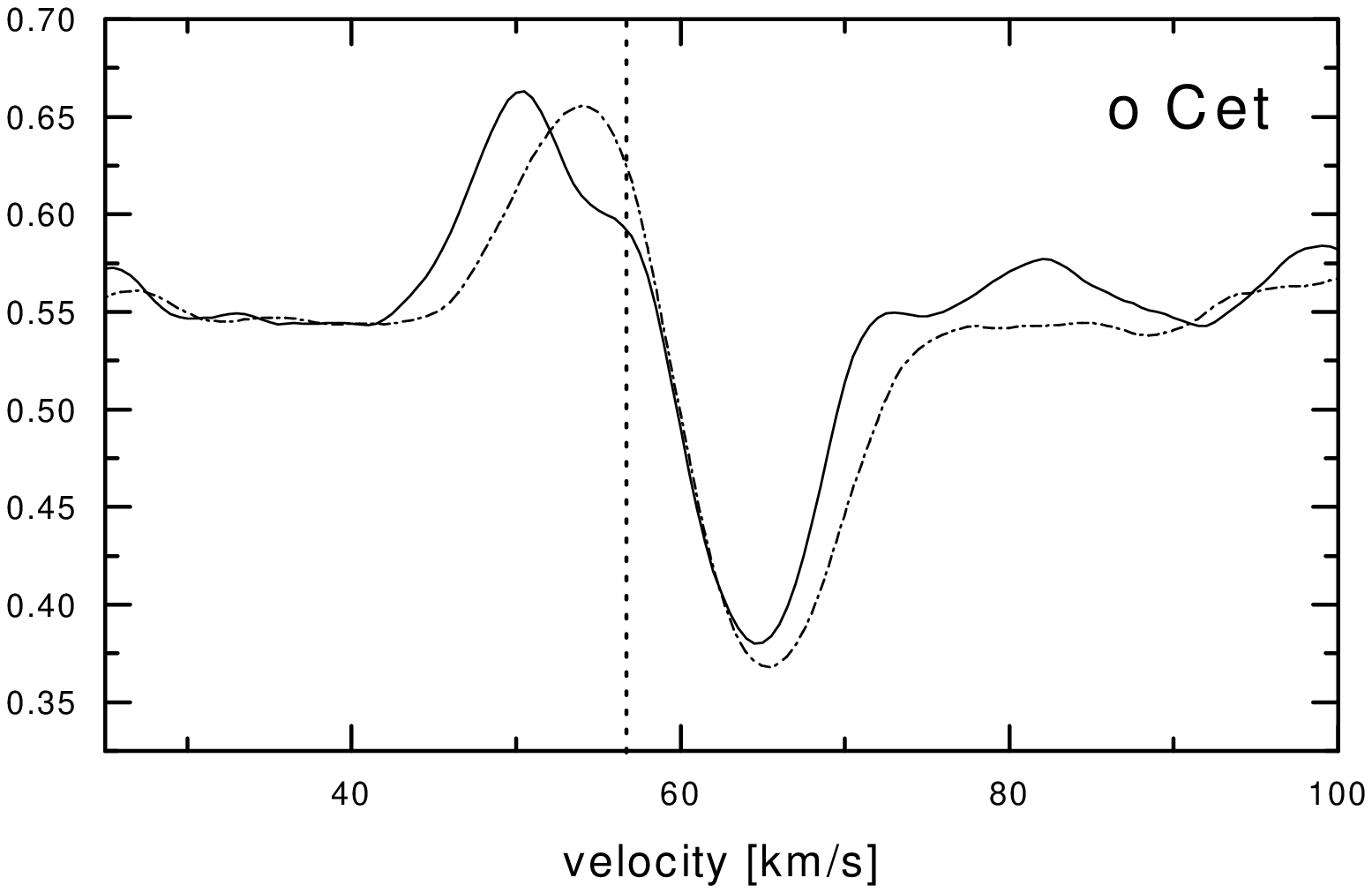}}
\end{center}
\caption{Averaged SiO line profiles from two observations of o\,Cet at
similar phase but widely separated in time. The solid line is the
profile from Feb 11 1985 ($\Phi =$ 0.82), while the dash-dotted line
marks the profile at Oct 21 1977 ($\Phi =$ 0.89). The dotted vertical
line indicates the center-of-mass velocity.}
\label{ocetsio}
\end{figure}

\begin{figure}
\begin{center}
\resizebox{6.5cm}{!}{\includegraphics{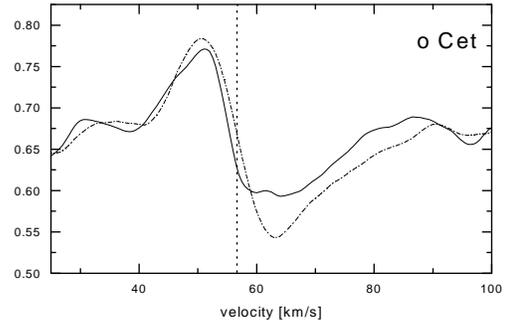}}
\end{center}
\caption{Same as Figure\,\ref{ocetsio} for OH.}
\label{ocetoh}
\end{figure}

Time series of averaged SiO, H$_2$O, HCl and OH lines of R\,Cas are plotted in
Figs.~\ref{rcassioh2o} and \ref{rcashcloh}, respectively. Typical line
profiles from the other three miras can be found in Figs.~\ref{ccygohh2o}
to \ref{ocetoh}.
No continuum is marked in the figures as the true continuum is
at a much higher level due to the almost continuous absorption by
weak H$_{2}$O lines (and similar molecules) in this spectral region. 
The line profiles shown are averaged line
profiles so that the range to both sides of the line can be used to
define some kind of pseudo continuum. In Figs.~\ref{rcassioh2o} to
\ref{rleosioh2o} spectra have been normalized to this pseudo continuum. The
center-of-mass velocity, for all objects from radio observations of
thermal CO, is marked in the figures (references on the center-of-mass
velocity as well as previously measured infrared and visual velocities
can be found in Hinkle et al.~\cite{HHR}, \cite{HSH84}, \cite{HLS97}).
Lines could be detected at all phases but vary considerably in shape
and strength.  A quick comparison of similar phases shows sometimes large
variation in the line profiles between the same phase in different
light cycles for the same star (e.g.~compare the spectra obtained at phase 0.61 and
0.62 in Fig.\,\ref{rcassioh2o}).  In the 4\,$\mu$m region R\,Cas 
is an extreme example of
cycle-to-cycle variations.  The other miras of our sample show some
differences between line profiles obtained from different light cycles
at similar phases but these differences are typically not as large as
in R\,Cas.  Note that observations
obtained at similar phase but several years apart can also have very
similar profiles (e.g.~Fig.\,\ref{ocetsio}).  This is suggestive of a strong aperiodicity as well
as a periodic variation in the spectra.  This is a common comment made
about visual spectra and some features (e.g. low excitation 2-0 CO
lines) in 2\,$\mu$m spectra but is not seen in the highly periodic
behavior of the great majority of lines in the 1.6-2.5 \,$\mu$m
region (e.g.\,Hinkle et al.\,\cite{HHR}).

Emission components are clearly visible in several observations. This
suggests a filling of the absorption with emission as an explanation for
the weakening of the SiO lines observed in low resolution spectra as
proposed by Aringer et al.~(\cite{AHWHJKW99}).  Multiple absorption
components with varying strength are an obvious characteristic of the
investigated lines in most of the spectra.  Unresolved multiple emission
components also appear.

For a given observation the four molecules investigated, OH, H$_2$O,
SiO and HCl, differ quite strongly in the shape of their line
profiles.  OH, as seen in the R\,Cas spectra, can show an emission line
while the other three molecules show an absorption line. Among the studied
oxygen rich miras differences of this type are largest for R\,Cas and
smallest for R\,Leo. In $\chi$\,Cyg OH and H$_2$O behave much more
similarly than in the other stars of our sample as illustrated in
Fig.\,\ref{ccygohh2o}.  Furthermore, the four molecules also differ in
their variability.  In the following we will now discuss the line
profile variations for each molecule separately.  We will offer an
interpretation of the complex line structures seen keeping in mind the
limitations of doing this without detailed model calculations.

If not explicitly stated velocities and
velocity shifts mentioned in the following are relative to the center-of-mass velocity of the
star.  The convention is used with negative velocity indicating outflow of 
material. 
The line profile figures have heliocentric velocity for 
the abscissa.

\subsection{OH}

\begin{figure}
\resizebox{\hsize}{!}{\includegraphics{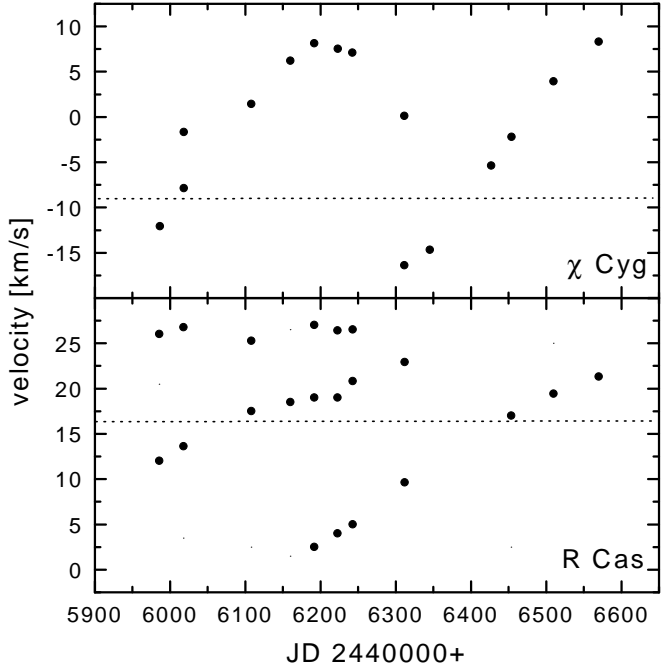}}
\caption[]{Velocities measured from absorption components of OH lines in $\chi$\,Cyg
(upper panel) and R\,Cas (lower panel). The dotted line indicates 
the center-of-mass velocity.}
\label{ohvel}
\end{figure}

Fig.\,\ref{ohvel} gives the resulting velocities derived from OH lines
for R\,Cas and $\chi$\,Cyg plotted against time.  Time is selected for
the abscissa rather than phase since
the variability does not always follow the photometric period.
Velocities were determined only for those absorption components of the line which
could be clearly identified and attributed to the line investigated. As
can be seen in Figs.~\ref{rcassioh2o} and \ref{rcashcloh}, other weak
absorption components can exist in the line profiles.  Obviously for
complex line profiles such as these the velocities are subject to
interpretation.  For instance, doubled absorption can result from an
emission line filling the center of a broad absorption line. 
We base the
results extracted from the averaged line profiles on similarities
found for CO lines of different excitation published before
(e.g.\,Hinkle et al.\,\cite{HHR}). For
$\chi$\,Cyg only the measurements between 1984 and 1986 have been
plotted to allow for the same time interval as R\,Cas.

For both stars we see a velocity curve resembling the velocity curves
found for CO first overtone high excitation lines in the 2.3\,$\mu$m
region (see
e.g.~Hinkle et al.~\cite{HHR}).  However, a closer inspection reveals
significant differences between the two stars and relative to the CO
velocity curve.  For R\,Cas we have a phase of line tripling between JD\,2446190 
and JD\,2446300 (see the individual line profiles in
Fig.\,\ref{rcashcloh}).  The first part of the velocity curve shows a layer
at rather constant infall velocity and a variable component changing
from outflow to infall.  During this time of infall, absorption is
becoming visible at an outflow velocity again. Surprisingly, the
temporal separation between observations of similar velocity is shorter than the
pulsation period of 430 days, in the one cycle examined it is $\sim$300 to 350 days. 
However, this behavior is clearly not related to the general pulsation of the 
star since the AAVSO light curve of R\,Cas at the time of the 
infrared spectroscopy
shows a cycle length of the typical 430 day photometric period.  The 
only notable deviation from the
normal light cycle is a very bright maximum preceding
the time these spectra were observed.  The first light maximum we
monitored has been a slightly weaker than usual. The absorption at
constant infall velocity is not visible in the second half of the
studied time interval and there is also no indication of a further line
doubling at the end of the second cycle around JD 2446600.

In $\chi$\,Cyg OH shows velocity 
changes with the period of the photometric variations,
i.e.~observations of similar velocity are separated by roughly one
period of the star.  Between the two cycles we have one observation with
absorption line doubling.  We also found line doubling on JD\,2446749,
but with a much smaller velocity difference. In addition, we note
that the averaged line profiles of $\chi$\,Cyg indicate further phases
of line doubling, but in these cases the second component is very weak
not allowing an accurate velocity determination.  

Beside the absorption components OH emission is visible at several phases
and can be quite strong. In the R\,Cas spectra an
OH emission component can be detected at all phases, but in several
spectra it is not obvious since the emission peak is
embedded in the absorption profile resulting in a weaker emission line that
does
not exceed the level of the continuum. The emission component starts at the
blue end of the line at a velocity relative to the center of mass 
of roughly -14.5\,km\,s$^{-1}$ (i.e.~+2\,km\,s$^{-1}$ heliocentric)
and
moves redwards. Around JD\,2446312 the emission component reaches
the center-of-mass velocity. Four months later a new emission line is
visible at the blue edge of the line again. It is not clear whether the
first emission line is still visible -- now at an infall velocity --
on JD\,2446510. On the other hand, the second emission line can already be
visible when the first one reaches the center-of-mass velocity.
Unfortunately, the coverage of the variability is not very good during
this phase.

We conclude that the line profile of OH in R\,Cas is
produced by the combination of one or two absorption lines, one outflow
and one infall component, which means the occurrence of line
doubling at some but not all phases. An emission line is running from
the blue to the red end of the red shifted line. The absorption lines
vary in strength and slightly in velocity during the cycle and from
cycle-to-cycle.  The emission component
can become quite strong, so that it becomes visible above the continuum
level. A weak emission component with an infall velocity seems to be
present in several spectra.  The line profiles of the other three miras
of our sample can be interpreted in the same way. In this way we can
also explain the unexpected line doubling in $\chi$\,Cyg on JD 2446006 
as the passage of an emission component through the line core.

\subsection{SiO \& HCl}

Representative line profiles for SiO and HCl in the   
R\,Cas, o\,Cet and R\,Leo spectra are illustrated in Figs.\,\ref{rcassioh2o},
\ref{ocetsio} and \ref{rleosioh2o}, respectively.  Strong emission
components of SiO and HCl are seen in several spectra 
of R\,Leo and o\,Cet, and also
in the spectra of $\chi$\,Cyg. In R\,Cas emission components are
clearly weaker. 

\begin{figure}
\resizebox{\hsize}{!}{\includegraphics{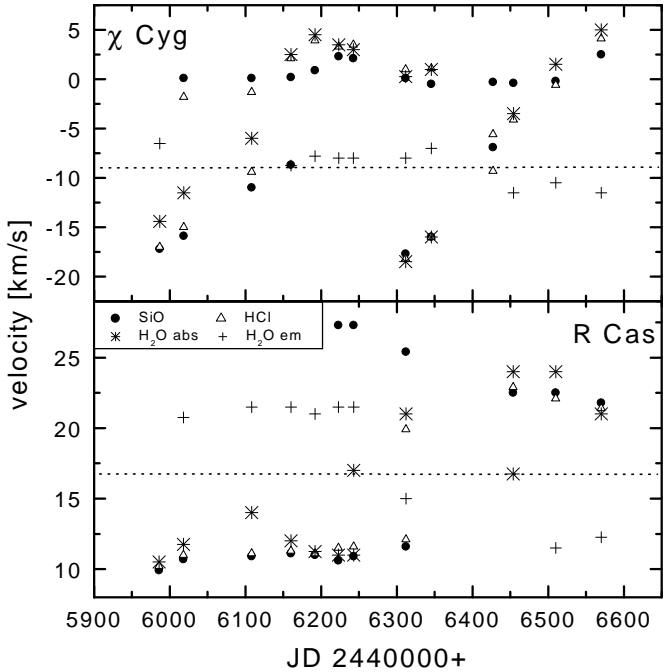}}
\caption[]{Heliocentric velocities measured from absorption components of SiO (filled circles), 
HCl (open triangles), water absorption (stars) and emission (crosses) lines in $\chi$\,Cyg
(upper panel) and R\,Cas (lower panel). The dotted line indicates the 
center-of-mass velocity.}
\label{siovel}
\end{figure}

The line profiles of these two molecules show similar and complex 
behavior with both emission and absorption.  As examples the velocity curves of
R\,Cas and $\chi$\,Cyg are discussed (Fig.\,\ref{siovel}). 
One absorption
component of the SiO and HCl lines
in $\chi$\,Cyg at a velocity of about +10\,km~s$^{-1}$ relative to the systemic velocity
of the star (about +0.5\,km\,s$^{-1}$ heliocentric) is visible in all SiO and HCl spectra of our data set,
including spectra obtained outside the time range plotted in
Fig.\,\ref{siovel}. In addition there is in $\chi$\,Cyg  a second absorption component
at several phases shifting its velocity 
towards that 
of the first component.  The period observed in these data
is about 320 days, conspicuously less than the photometric period.
Note that the velocity curves of the two cycles of $\chi$\,Cyg observed are 
not parallel. We will discuss this difference in more detail in the next section.

R\,Cas has quite different velocity
variations compared to the other miras studied.  From 
Figure\,\ref{rcassioh2o} we see that from JD\,2445987 to
JD\,2446192 the outflow component of the SiO lines is dominant in the
spectrum, while from JD\,2446454 to JD\,2446570 the infall component is
clearly stronger.  Yet the absorption components have similar depths at
similar phases.  In the spectra obtained in 1986 (JD\,2446430 through
JD\,2446800) the outflow
component may be filled by an emission line which is becoming visible
in the JD 2446570 spectrum.

A strong emission component is clearly visible in several spectra of
R\,Leo (see Fig.\,\ref{rleosioh2o}), o\,Cet and $\chi$\,Cyg (Fig.\,\ref{ccygohh2o}). 
The R\,Leo data set is an example of spectra where the line profiles 
can be interpreted as a combination of an absorption line and an emission
line, with the absorption and emission having different phase 
dependent velocity behavior. 
On the other hand, in R\,Cas
an additional weak emission component at the systemic velocity is seen as
an obvious asymmetry in the absorption line.
In the R Cas spectra near maximum light, and especially the spectrum of
JD 2446192, the strong absorption component of SiO and HCl
matches the velocity of the strong (for R\,Cas) OH emission line.

As the variability cycle of the 4\,$\mu$m molecular features does not
always follow the photometric period it is very difficult to compare the line
profiles of different miras at similar visual phases directly. However,
basic characteristics, namely outflow and infall components, 
intensity variations, and the occurrence of emission at some phases, are found in
all miras of our sample.  The similarity between the SiO line profiles
and the HCl line profiles
is a common feature of all stars. In general,
line profile and variability is similar in R\,Leo, o\,Cet and
$\chi$\,Cyg. R\,Cas shows a different, more complex behavior.

\subsection{H$_2$O}

The H$_2$O line profiles do not follow the variations of either the OH
or the SiO and HCl line profiles. This is illustrated by comparison of
H$_2$O line profiles with those of other molecules in
Figs.\,\ref{rcassioh2o}, \ref{ccygohh2o} and \ref{rleosioh2o}.  The
dissimilarity of profiles indicates that, at least at some phases, the
H$_2$O lines probe a very different atmospheric zone than the lines of the
other molecules seen in the 4\,$\mu$m spectra.  In addition, we note
that the H$_2$O lines in $\chi$\,Cyg are weak in agreement with the
S-type spectral class of this star (Fig.\,\ref{ccygohh2o}).  Emission
components, which can be seen in other molecular lines, are even more
prominent in H$_2$O line profiles.

Velocities derived from absorption and emission components of the
H$_2$O line profiles of R\,Cas and $\chi$\,Cyg are plotted in
Fig.\,\ref{siovel} together with the results from SiO and HCl. The
$\chi$\,Cyg H$_2$O absorption lines show a velocity variation similar
to the other three molecules monitored in this stars spectrum. The
velocity curve is slightly offset from the SiO and HCl variations,
mimicing the velocities of the OH and CO lines but with smaller
amplitude. In the one case
where the CO and H$_2$O velocity significantly differ the cause is
probably a blend of the emission and absorption line profiles.  The
emission component stays almost constant in velocity close to the
center of mass velocity throughout the entire light cycle.

R\,Cas velocities (lower panel of Fig.\,\ref{siovel}) again have quite
different phase dependent behavior than those of $\chi$\,Cyg.  One
H$_2$O absorption component shows a S-shaped discontinuous velocity
curve apparently spanning the entire time of observations, i.e.~at
least two light cycles!  Additional absorption components resemble the
velocity behavior of SiO and HCl around JD 2446200. The emission
component seen in $\chi$\,Cyg close to the center-of-mass velocity is
not found.  Emission is instead seen at an infall velocity first and
switches to an outflow velocity shortly after the visual light maximum
at JD2446200.  The overall behavior of the 4\,$\mu$m spectral lines in
R\,Cas is outstanding and will be discussed below.

\section{Mira Dynamics}
Monitoring of the velocity variations of spectral lines provides a
powerful tool for studying the physical characteristics and 
dynamics of the stellar atmosphere. In this section we summarize 
observed velocity variations for the two miras with the most
complete time series, $\chi$\,Cyg and R\,Cas. On this basis we
discuss the dynamics of the outer layers of an AGB star's atmosphere.

There are two basic features of the 4\,$\mu$m spectra that need to 
be explained by any model:  
the occurrence of an emission component
and the deviation of the variability from the photometric period. However,
we stress that even a qualitative description of these phenomena has to
be limited to some basic statements due to the complex, 
not fully understood structure of AGB atmospheres.  
It can be difficult to discern observationally the movement of 
a layer and a change in the optical
depth since they can produce interfering results. The emission component may have
its origin in the atmospheric dynamics or in a geometrical effect.
We will discuss the absorption components first and come back to
the emission lines at the end of this section.

\subsection{$\chi$\,Cyg}
The $\chi$\,Cyg spectra described
above provide the most extensive and well sampled time series of
4\,$\mu$m spectra of any star in our sample.  $\chi$\,Cyg also has good
time series spectra in the H-band and K-band which cover the first and
second overtone lines of CO.  We will summarize the velocity data for
this star as an example of the overall relation between the dynamics of
atmospheric layers as probed by different molecular features in the
infrared.

The second overtone CO lines at 1.6\,$\mu$m are thought to provide
information about the kinematics deep inside the atmosphere, i.e.~close
to the pulsation driving zone (Hinkle \cite{Hinkle78}). The
corresponding velocity curve is S-shaped with line doubling occurring
around maximum in all miras.  The line doubling is seen only in miras,
not in the lower amplitude semiregular variables.  In miras the
center-of-mass velocity is crossed by the velocity curve typically
around phase 0.4. At that phase no line doubling occurs 
in these stars.  A similar behavior is shown by the
high excitation first overtone CO lines found in the 2.3\,$\mu$m
region. For $\chi$\,Cyg an extensive time series of these lines has
been published already (Hinkle et al.~\cite{HHR}). For the present
investigation a new set of 1.6\,$\mu$m spectra was taken in order to
compare the 4\,$\mu$m results with $simultaneous$ data from CO lines.
The 1.6\,$\mu$m spectra were obtained with the same instrumental
setup as the earlier measurements published in Hinkle et al.~(\cite{HHR}).

\begin{figure*}
\resizebox{12cm}{!}{\includegraphics{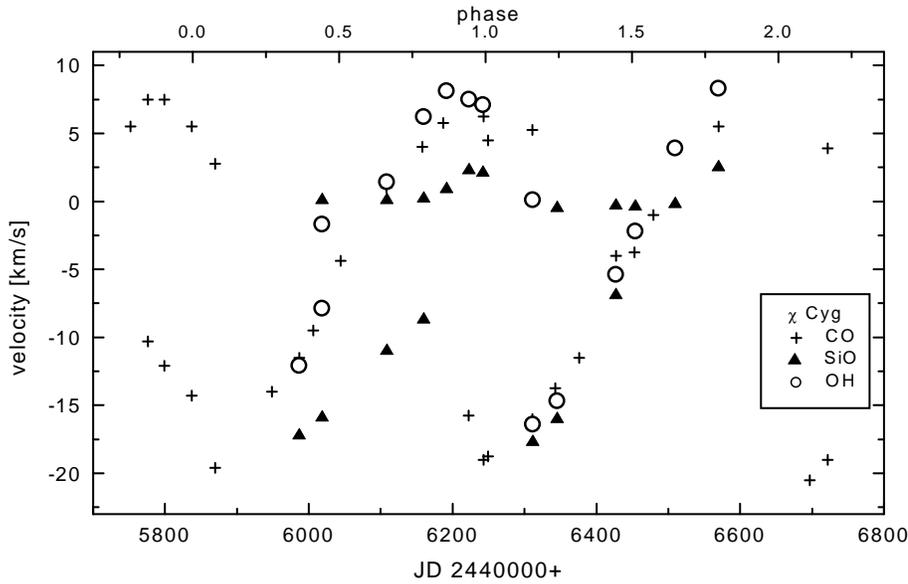}}
\hfill
\parbox[b]{55mm}{
\caption{Velocities of different molecular lines versus time and
visual phase for $\chi$\,Cyg. The visual phase given
at the top axis is relative to the visual light maximum at
JD 2445838.
Crosses mark velocities from CO lines, triangles
SiO velocities and open circles OH velocities, respectively.}
\label{allvelchicyg}}
\end{figure*}

Figure\,\ref{allvelchicyg} shows the velocities of the different
molecular species versus time for the mira $\chi$\,Cyg.
Visual phase
relative to the first maximum observed (phase 0.0) has been added as
label to the top axis.  As demonstrated previously (Hinkle et
al.~\cite{HHR}) the CO lines vary with the photometric period. 
Line doubling is visible around maximum.  There is
no difference between the CO velocity curve presented here and those
obtained earlier by Hinkle et al.~(\cite{HHR}). 
The 4\,$\mu$m OH lines in $\chi$\,Cyg follow the behavior 
of the CO lines.  
In general the H$_2$O absorption line velocities for $\chi$\,Cyg follow 
the CO velocity curve.

The velocity curve defined by the SiO lines is
clearly different from the CO line curve. 
As mentioned above SiO has a more or less
constant velocity component at an infall velocity of about 
10\,km\,s$^{-1}$. There is
no feature in the CO or OH profiles that could be correlated with the
first absorption component of SiO at constant velocity.
A smaller velocity amplitude is seen in SiO (and also in
HCl, see Fig.\,\ref{siovel}) lines.

SiO also has a second absorption component that 
shows strong variability in velocity.
But this variation is not always quite with the 
stellar pulsation. 
During the first cycle observed for $\chi$\,Cyg the line forming region
of SiO decelerates less than the CO and OH line forming region, assuming
that the lines of a certain species originate in the same layer at all phases. 
During this cycle the braking effect is smaller for this
layer. In
the second light cycle SiO has a velocity component that 
follows the CO velocity curve. 

A possible explanation is that mass loss is not continuous, with
material ejected in shells but not during every light cycle
(e.g.~Winters et al.\,\cite{Winter00}).  Several observations in the
radio range and of post-AGB objects support this description of the
mass loss. If so, material would be aggregated at some distance from
the star before it is ejected. A mass shell reaching this distance during
its pulsation will be then confronted with more or less material,
i.e.~with an environment with varying density from cycle-to-cycle. If
there is less material, the layer or a moving wave can go faster and further
away from the star and in this
way again would
feed the mass loss. The problem with the time between phases of similar
velocity being shorter than the photometric period can in that way be
explained. The layers traced by the SiO (and also by HCl)
are not varying on a shorter
time scale but with a different acceleration during some cycles. After
several cycles we can again have similar line profiles as 
observed in o\,Cet and $\chi$\,Cyg. 

\subsection{R\,Cas}

The mira R\,Cas shows a behavior remarkably different from $\chi$\,Cyg,
especially corresponding the velocities of OH and SiO lines. No simultaneous CO
observations are available for this star.  Comparison with older data
(Hinkle et al.~\cite{HSH84}) shows that the velocity curve derived from
the OH lines has a flatter slope than the CO velocity curve. The
maximum expansion velocity observed in OH is smaller than the one
observed in the CO lines.

\begin{figure}
\resizebox{\hsize}{!}{\includegraphics{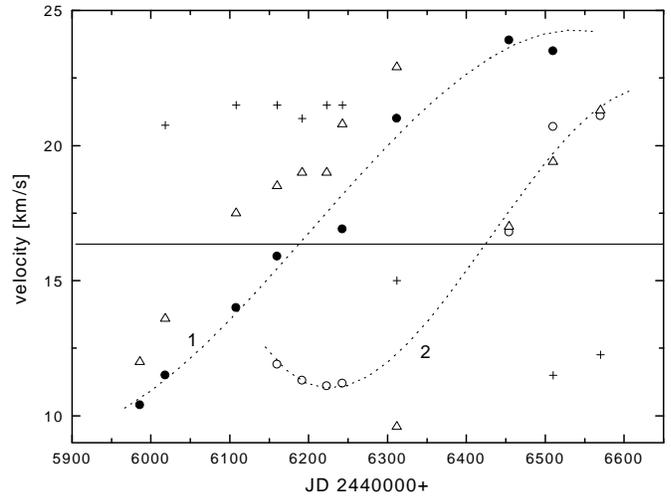}}
\caption{H$_2$O and OH velocities of R\,Cas versus time. Filled and open circles mark
the two H$_2$O absorption components, crosses indicate the velocities of water emission
lines, and open triangles denote OH absorption lines. Two velocity curves, marked '1' and
'2', have been drawn through the H$_2$O data points. The horizontal line indicates the
center of mass velocity. See text for more details.}
\label{rcasdynam}
\end{figure}

The H$_2$O lines in R\,Cas are the only ones found that indicate the
presence of (periodic) variations in the outer layers of the atmosphere with a
time scale longer than the period indicated by the visual light
change.  In Fig.\,\ref{rcasdynam} we show that the different H$_2$O
absorption components can be grouped to two nicely shaped velocity
curves by two hand drawn fits through the data points. The curves are 
numbered '1' and
'2', respectively. 
Curve '1' as noted above covers a time scale exceeding the photometric
period, while the variation represented by curve '2' happens on a time scale similar to
the main period. Curve '2' also shows the same relation between
velocity and light change found e.g.~for high excitation CO lines,
i.e.~minimum velocity at light maximum.  However, no line doubling
of the same type as in the CO lines is
observed for curve '2', as any second absorption component fits on
curve '1'.

Fig.\,\ref{rcasdynam} shows the velocities of the OH lines as well. It
can be seen that they form a velocity curve of similar shape as the
H$_2$O lines but with an offset.  The first part of the
OH velocity curve shows also the
same velocity {\it range} as the H$_2$O lines while the second velocity
curve found from the OH lines has a larger range in velocity. This may
indicate that our time series does not cover the whole length of this
long period change, so that the time scale may be even larger than two times
the main period.

The line doubling of the H$_2$O lines suggests that during some
phases we see line forming layers at two completely different locations
in the stellar atmosphere.  The similarity in the behavior of curve '2' 
and the high excitation CO lines implies that curve '2' represents a layer close to the
pulsation driving zone. It is interesting to note that we did not find any
absorption components that would mark this layer around JD2446000.

Based on these results we propose the following scenario occurring in
R\,Cas: Around JD2445800, i.e.~at the light maximum preceding our time
series, a cycle of velocity variations started that 
lasted much longer than a typical cycle. A similar model as for the SiO
line forming layer in $\chi$\,Cyg may therefore be assumed, i.e.\,that 
the variation marked as
curve '1' in Fig.\,\ref{rcasdynam} lasted longer than one cycle as material
may have been moved further away from the star and needed a longer time to
return to its original position. 

Note that our data do not give any
indication whether this behavior is occurring with some periodicity or not.
The CO lines, observed at a different time, did not show such strange
behavior. 
In this context it is
interesting to note that a very bright maximum is preceding this phase
of strange velocity behavior. Such bright maxima occur from time to
time with no obvious regularity in R\,Cas (Mattei \cite{Mattei00}). It
seems quite probable that these two phenomena are related to each other. 
At
the next light maximum again a velocity curve starts, this time its
period is following the photometric changes. 
Therefore, only in some cycles such strange line profiles
may occur. 

However, as the behavior of the SiO lines in R\,Cas shows, this
scenario cannot explain the dynamics of the outer atmosphere
completely. The SiO lines obviously vary on a time scale longer than
the photometric period, but from our data there is no indication for
any periodic variation.

\subsection{Emission lines}
Emission can in principle be produced by a temperature inversion, as
it would be provided by a shock front or a hot chromospheric layer, or by a
geometric effect in an extended spherical object (e.g.~Hinkle et
al.\,\cite{Kenh2}).  Both mechanisms are possible in miras.  The
occurrence of shock fronts is well known not only from optical spectra
but also from the velocity curves derived from second overtone CO lines
in the near infrared (e.g.~Hinkle et al.\,\cite{HHR}).  In the optical
range emission of hydrogen and several metal lines is triggered by the
passage of shock fronts. Chromospheric layers have been proposed from
model calculations by Willson \& Bowen (\cite{WB}), but no final
observational proof has been given for their existence in oxygen
rich AGB stars.  Finally,
atmospheres of pulsating AGB stars are definitely widely extended and
can show a shell like structure in the outer parts (e.g.~H\"ofner
\cite{Hoefner99}).

We note that the emission component of some molecules changes 
velocity during the
time series but other emission components have a near constant 
velocity.  This suggests two separate places for the emission.
In R\,Cas we observe an emission component in the OH lines
changing from a high outflow velocity to the center-of-mass velocity.
If we attribute the emission to a certain moving layer, we would see a
braking of the movement until it comes to a stop relative to the center
of the star. Such a behavior of the emission has also been seen in
emission lines of the Balmer series in miras (e.g.~Merrill
\cite{Merrill46}).  However, the emission seen in the H$_2$O lines,
most notably in $\chi$\,Cyg, is constant in velocity.  We propose
that this emission is produced in the outer layers of the extended atmosphere.  Tsuji
(\cite{T00}) makes similar arguments for emission seen from water 
in the inner circumstellar shell.  
The extended atmosphere may also explain emission seen in molecules
other than H$_2$O in  SRV's where 
there is no indication for a shock front from second
overtone CO lines or hydrogen lines (see below). 

\section{SRVs}

Semiregular variables are generally believed to have a lower pulsation
amplitude than the miras and as a result SRV spectra are expected to be
less affected by dynamic phenomena.  To study the influence of
stellar dynamics on the 4\,$\mu$m spectra of SRVs, we examine in
this section representative SR spectra from different period and
amplitude regimes.  Short period and low amplitude SRVs typically have
small mass loss rates and are therefore difficult to detect in
thermal radio CO lines (e.g.~\mbox{Kerschbaum} \& Olofsson
\cite{KO99}). As a result the most reliable information on the
center-of-mass velocity for long period variables is not available for
these stars.  The discussion of velocities relative to the center 
of mass is therefore limited to the subset of 
stars where an accurate systemic velocity has been
measured.

The SRV observations reported here cover only a single epoch.  Due to
the lack of reliable photometric data a rough determination of the
actual visual phase of the variables at the time of the observation was
possible only for three SRVs of our sample.  As a consequence we will
focus on W\,Cyg and RU\,Cyg, for which both velocity and phase
determination are quite reliable.

\begin{figure}
\begin{center}
\resizebox{6.5cm}{!}{\includegraphics{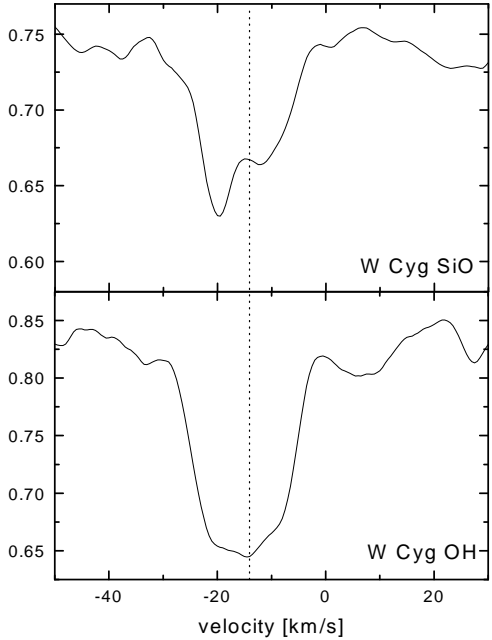}}
\end{center}
\caption{SiO (upper panel) and OH (lower panel) line profiles of W Cyg.
The dotted line marks the systemic velocity.}
\label{wcyg}
\end{figure}

\begin{figure}
\begin{center}
\resizebox{6.5cm}{!}{\includegraphics{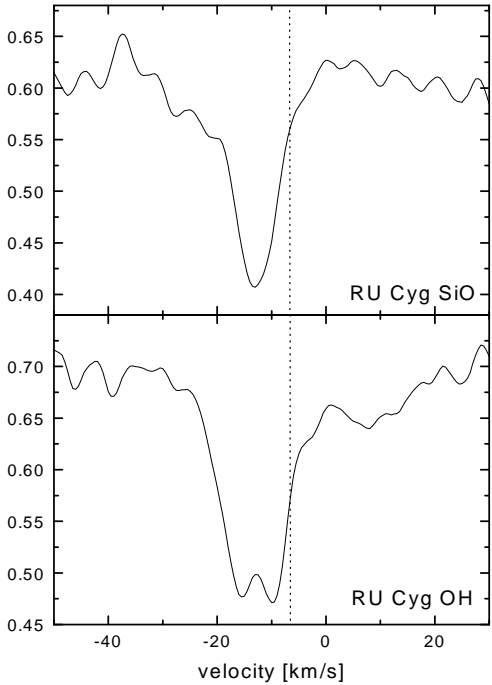}}
\end{center}
\caption{SiO (upper panel) and OH (lower panel) line profiles of RU Cyg.
The dotted line marks the systemic velocity.}
\label{rucyg}
\end{figure}

Figs.\,\ref{wcyg} and \ref{rucyg} show the SiO and OH line profile of
the semiregular variables W\,Cyg and RU\,Cyg, respectively.  The SiO
lines in W\,Cyg show an absorption component on each side of the
systemic velocity with the outflow component dominating. Outflow occurs
with a velocity of 6\,km\,s$^{-1}$. In contrast the second overtone CO lines show only
outflow velocities over the whole light cycle (Hinkle et
al.~\cite{HLS97}).  One of the CO second overtone spectra has been
obtained only 2 days before the 4\,$\mu$m observation and gives a
heliocentric velocity of $-$20.0\,km\,s$^{-1}$, i.e.~an outflow velocity of about
6\,km\,s$^{-1}$ the same as found for the SiO lines. Compared with
miras at similar phases (e.g.~R\,Leo, line profile at the bottom of
Fig.\,\ref{rleosioh2o}) the SiO lines in W\,Cyg are weaker. It should be remarked that 
the SiO lines in W\,Cyg are the weakest among all the SR spectra in our
sample.  Comparison with miras is of course difficult due to
nonperiodicities.  The average OH line profile of W\,Cyg (bottom panel
of Fig.~\ref{wcyg}) is broad and slightly asymmetric towards the
outflow part.  As for miras, one possibility is that the doubling of
the absorption is produced by an emission component filling in the
absorption line, although no indication for a shock front is given in
the second overtone CO line velocity curve for W\,Cyg.   However, the
emission is strongly supported by the H$_2$O profile which shows
emission above the continuum with weak absorption on each side.  The
emission is  at $-$20\,km\,s$^{-1}$ heliocentric.

The SiO and OH profiles in RU\,Cyg are quite different from those of
W\,Cyg.  A double structure is visible in the OH lines but not in the
SiO lines.  The spectrum indicates only outflow of matter. As for
W\,Cyg a spectrum with second overtone lines of CO has been obtained
almost simultaneously (Hinkle et al.~\cite{HLS97}). The CO lines
indicate an outflow, too, with a similar velocity. The
profile could of course result from an emission/absorption blend.  The
$\nu_3$ lines of water can not be detected in RU\,Cyg.

The other five SRVs show slightly asymmetric SiO lines 
with the positive velocity side steeper than the negative velocity
side.  None of these objects shows any indication of 
line doubling or emission components in the line core.  SiO lines
are more narrow in g\,Her, R\,Lyr and RZ\,Ari, i.e.~in the stars with
periods below 100 days, than in the other SRVs.  The OH lines look
similar to the SiO lines in these three SRVs.  For g\,Her a reliable systemic
velocity is available (Kerschbaum \& Olofsson \cite{KO99}).  Neither the
SiO lines nor the OH lines show a significant shift relative to this
systemic velocity.  HCl lines have been detected in all SRVs except for
W Cyg. They are much weaker in these stars than in miras.  

H$_2$O is 
detectable in three of the SR stars, RX Boo, g Her, and SW Vir.
In all three it is seen in emission.  In SW Vir the H$_2$O line has an
inverse P Cyg
shape with absorption at $-$18.4\,km\,s$^{-1}$ and emission at 
$-$11.3\,km\,s$^{-1}$ heliocentric.  The H$_2$O velocity for RX Boo is 
$-$11.3\,km\,s$^{-1}$ and +6.8\,km\,s$^{-1}$ for g Her (heliocentric).
Comparison with center of mass velocities from radio observations
shows that the water emission occurs always close to but not necessarily
exactly at the systemic velocity of the star. The largest difference was
found for g\,Her (CO radio velocity: +1.3\,km\,s$^{-1}$).

\section{Conclusions}
\subsection{Explaining low resolution spectra}
Considerable variations in the strengths of SiO bands during a
light cycle have been observed in low resolution spectra of AGB
stars. In our high resolution observations the lines 
are convincingly present at all
phases, showing a presence of SiO in the atmosphere throughout the 
light cycle. 
Therefore it is obvious that the simplest proposals, e.g. the dissociation of 
a large fraction of molecules near maximum light, do not explain 
the spectrum.  This has been
suggested to explain the extreme weakening of the SiO bands near light
maximum observed in low resolution spectra (e.g.\,Rinsland \& Wing
\cite{RW82}). The identification of emission components in
these lines suggests the blending and cancellation of absorption and 
emission as the
explanation for the weakening (Bessell et al.~\cite{BSW96}, Aringer et al.~\cite{AHWHJKW99}). 
It is interesting to note that Aringer et al.~(\cite{AKHH00}) found a
significant emission in 4\,$\mu$m range of ISO SWS spectra of R\,Cas, which they
attributed to SiO, while we found only a rather inconspicuous emission
component in our high resolution spectra.  This discrepancy suggests
strong cycle-to-cycle variations which warrant further investigation.

\subsection{Dynamics of miras}
A surprise in the 4\,$\mu$m spectra is the large degree of combination 
of velocity variations that arise close to the  
pulsation driving zone with grossly non-periodic
variations arising in extended atmospheric layers.  
In general the continuous opacity in the 4\,$\mu$m region 
is significantly higher than in the 1.6\,$\mu$m range.
As a consequence there is a tendency for lines to originate
further out.  
Therefore many molecular features seen at 4\,$\mu$m monitor stellar
variability in the outer parts of the atmosphere, especially many
SiO and H$_2$O lines. 
A similar increase in the continuous opacity
toward the blue is well documented and seen in the appearance of the
visual spectrum.  The line profile variations 
seen at 4\,$\mu$m appear similar to those found
for \ion{K}{i} and \ion{Fe}{i} lines around 8000\,{\AA} by Gillet et
al.~(\cite{GMBF85}). Part of the periodic and
non-periodic variations seen in the 4\,$\mu$m region could also be due
to a deviation from spherical symmetry.

The four stars in our sample show both strong similarities and
differences in the 4\,$\mu$m lines. Especially for R\,Cas we find a quite
different behavior for most of the lines.
We conclude that, while the lower regions of the atmosphere
where pulsation originates are very similar in all the miras, the outer
layers of the mira atmosphere are much more distinctive.  This could be
the result of evolutionary state along the AGB as might be suggested by
the range of periods. Other possibilities are differences in the 
fundamental properties like mass,
metallicity, or C/O. Or the behavior is unique for each star resulting from spatial
inhomgeneities in the atmosphere.  Our sample is too small to differentiate
between these possibilities.  However we note that R\,Leo and o\,Cet
have a similar period and similar C/O (based on the water and CO line
strengths) and have more similarity in their 4\,$\mu$m spectra than
with the other two miras. $\chi$\,Cyg, although having a similar period
to R\,Cas, is an S-type star so differences between the line profiles
in these stars are expected to be produced by the different atmospheric structure
which is related to chemical composition.

Fig.\,\ref{allvelchicyg}
shows, that the lines can be a tracer of
the pulsation, like those in the 1.6\,$\mu$m spectrum, but that in 
some cycles irregular behavior occurs.  Periodicities in the
outer layers different from the photometric period have been suggested from
dynamical model atmospheres (see e.g.~Fig.\,1 in H\"ofner \& Dorfi
\cite{HD97}, and Hofmann et al.~\cite{HSW98}). The observational
findings can be qualitatively understood in the framework of these dynamical
models of AGB star atmospheres. During some cycles, which seem to be marked by
a brighter visual maximum, part of the outward moving layers decelerates less than
in other cycles due to a smaller density in the outer layers. This density variation
may be related to a non continuous mass loss. 
Such deviations are not monitored by lines formed deeper inside the atmosphere.
It is therefore an effect only occuring in the outer parts of the star. This means
that this behavior is not related to a different strength of the pulsation.

In this context it is interesting to note, that Rinsland \& Wing
(\cite{RW82}) made two low resolution 4\,$\mu$m spectra of each of the
miras R\,Cyg, $\chi$\,Cyg and R\,Cas separated by almost exactly one
light cycle. In two miras, R\,Cas and $\chi$\,Cyg, they detected a
clear difference in the overall strength of the SiO (2,0) and (3,1)
bands between the two observations made at similar phase. This is in
good agreement with the findings discussed above.

\subsection{SRVs}
Small amplitude variables like the SRVs of our sample are thought
to be more compact than miras. Variations of CO lines at 1.6\,$\mu$m
(Hinkle et al.\,\cite{HLS97}, Lebzelter\,\cite{Lebzelter99})
indicate that SRVs vary with a much smaller velocity amplitude 
and that shock fronts are very weak or do not occur
at all. In agreement with this
finding, the SRVs of our sample
do not show such extreme emission components as
found in miras. 
The line profiles of most SRVs show a similar shape and
strength. This indicates that their outer atmospheric layers show a
similar structure. Due to the lack of time series for these stars no
study on their individual variability is possible.  

As our main result for SRVs we note that both the occurence of emission
components as well as the asymmetries and the shifts relative to the
center of mass velocity clearly indicate that velocity fields have
to be included into a model atmosphere to reproduce the results properly.
Hydrostatic models cannot be used for high resolution spectra of these
stars. However, the fact that the effects of the stellar pulsation are
still rather small in SRVs compared to miras makes these stars excellent
starting points for adapting dynamical models to the observations.

\section{Outlook}
The aim of this paper was to present and to discuss qualitatively the
observational results of our high resolution spectroscopic monitoring
of AGB variables. In a forthcoming paper we will present synthetic 
spectra based on dynamic model atmospheres in order to model the observations
presented here.

However, an obvious explanation of the non-regular variations is
inhomogeneities in the upper atmosphere.  These 
inhomogeneities could be in the form of
filaments or clouds in the extended
stellar atmosphere.  
Unique filament structure would explain non-cyclic behavior 
and the strong spectral identity of individual stars.
There is considerable evidence from maser observations of such
condensations 
in the outer atmosphere.  The SiO maser, which is probably formed in 
the same atmospheric region as some of the 4\,$\mu$m lines, is
clearly formed in condensations (Diamond et al. \cite{d94}).
Large scale inhomogeneities are indicated by maps and images of AGB
circumstellar shells. A further indication for inhomogeneities in
the atmospheres of AGB stars come from polarimetry measurements.
As shown from model calculation (Harrington \cite{Harrington69})
observational results can be understood if it is assumed that
temperature variations occur over the stellar surface. This
model has been applied successfully to observations of
o Cet by Tomaszewski et al.~(\cite{TLMC80}). 
Such filament structures currently exceed foreseeable modeling 
capabilities, but it might be possible to include their effects 
in future dynamical model atmospheres.  

\begin{acknowledgements} 
We wish to thank Dr.\,Susanne H\"ofner for fruitful discussions on her model
atmospheres.  This work was supported by Austrian Science Fund Projects
S7308-AST and P14365-PHY.  
K.H.H. would like to thank the 
NOAO director Dr.\,Sidney Wolff for funding two visits to the Institut
f\"ur Astronomie which made this research possible.  K.H.H. would also like to
acknowledge the hospitality of Drs. Michel Breger and Josef Hron at
the Institut f\"ur Astronomie in Vienna.
This research made use of the SIMBAD
database operated by CDS in Strasbourg, France, and NASA's Astrophysics
Data System Bibliographic Services.  In this research, we
have used, and acknowledge with thanks, data from the AAVSO
International Database, based on observations submitted to the AAVSO by
variable star observers worldwide.  
\end{acknowledgements}

\end{document}